\documentclass[aps,prd,showpacs,amsmath,amssymb]{revtex4}
\usepackage{graphicx}
\usepackage{amsfonts,bm}
\usepackage[english]{babel}
\usepackage{color}

\def\be{\begin{equation}}
\def\te{\end{equation}}
\def\ee{\end{equation}}
\def\ba{\begin{eqnarray}}
\def\bea{\begin{eqnarray}}
\def\nn{\nonumber\\}
\def\tea{\end{eqnarray}}
\def\ea{\end{eqnarray}}
\def\eea{\end{eqnarray}}

\begin{document}

\title{A hydrodynamic approach to QGP instabilities}

\author{E. Calzetta}
\email{calzetta@df.uba.ar}
\author{J. Peralta-Ramos}
\email{jperalta@df.uba.ar}
\affiliation{Departamento de F\'isica, Facultad de Ciencias Exactas y Naturales, Universidad de Buenos Aires and IFIBA, CONICET, Cuidad Universitaria, Buenos Aires 1428, Argentina}


\begin{abstract}
We show that the usual linear analysis of QGP Weibel instabilities based on the Maxwell-Boltzmann equation may be reproduced in a purely hydrodynamic model. The latter is derived by the Entropy Production Variational Method from a transport equation including collisions, and can describe highly nonequilibrium flow. We find that, as expected, collisions slow down the growth of Weibel instabilities. Finally, we discuss the strong momentum anisotropy limit. 
\end{abstract}
\maketitle

\section{Introduction}

The hot and dense fireball of nuclear matter created in heavy ion collisions at the Relativistic Heavy Ion Collider and the Large Hadron Collider behaves as an almost perfect fluid, with a viscosity-to-entropy ratio not far from the lower  bound $\eta/s = 1/(4\pi)$ derived from the Anti de Sitter/Conformal Field Theory correspondence \cite{rev}. This implies that, in spite of the rapid longitudinal expansion, which limits the effectiveness of particle collision in equilibrating the system, the fireball isotropizes extremely fast with characteristic times $\lesssim 1$ fm/c. 

Mr\'owczy\'nski was the first to show that there exists an instability to chromomagnetic fluctuations in the Quark-Gluon Plasma (QGP) for wavevectors tranverse to the chromomagnetic field \cite{mro1,mro2} (see also \cite{mro3,mro4,mro5}). These are the analog of the Weibel instabilities in a electromagnetic (EM) plasma \cite{weibel} and occur if the one-particle distribution is anisotropic in momentum space (see also \cite{schlick,ach1,ach2}). This type of distribution function is relevant in heavy ion collisions because the longitudinal expansion of the fireball causes the system to become much colder in the longitudinal direction than in the transverse ones, implying 
$\left\langle p_L^2 \right\rangle \ll \left\langle p_T^2 \right\rangle$ in the local rest frame, where $p_{L,T}$ is the particle momentum in the longitudinal and transverse direction.

The transverse instabilities are found to have a strong impact on the fireball's evolution due to the large momentum anisotropy present at early times, speeding up the isotropization and equilibration process. Basically, the instability causes the soft sector of the magnetic field to become rapidly amplified, which leads to large-angle scattering of hard particles thus leading to faster isotropization and thermalization. Therefore, the thermalization process in the QGP is not controlled solely by particle collisions, but the role of collective effects (such as Weibel instabilities) must be taken into account as well.

The chromo-Weibel instability has been extensively studied \cite{mro1,mro2,mro3,mro4,mro5,schenke-coll,
dominik,rebhan,dumitru,dumitru2,attems1,attems2,ven,arnold,rom,rom2,rom3,ipp,moore}, both analytically in linear response and numerically in the fully non-linear case, either within the Hard-Thermal Loop framework or directly from the non-linear Vlasov equations for anisotropic plasmas (which go beyond the HTL approximation, see e.g. \cite{manuelrev} and references therein). One of the most important results that comes out from these studies is that the non-linear gauge self-interactions slow down the growth rate of the Weibel instability, an effect that could be studied with a suitable generalization to the non-Abelian case of the formalism presented in this work. 

More directly connected to our study is the work by Schenke and coworkers \cite{schenke-coll}, in which the authors study stable and unstable modes as obtained from the Boltzmann-Vlasov equation for non-Abelian fields (but in the effective Abelian approximation) with a Bhatnagar-Gross-Krook collision operator. We shall compare our results to those of Ref. \cite{schenke-coll} along the way. 

The usual kinetic theory approach to study plasma instabilities combines a collisionless transport equation for a one-particle distribution function with the Maxwell equations for the EM field, see e.g. \cite{ichi,krall}. In the background the distribution function is position independent and the Maxwell field vanishes (see however Ref. \cite{dominik} where initial fluctuations of the currents are taken into account). The linearized equations are a simple transport equation for the perturbation of the distribution function with a source which depends on the linearized Maxwell field. This equation may be solved exactly, leading to an integro-differential equation for the Maxwell field. Since the background is homogeneous, the perturbations may be Fourier analyzed, whereby a dispersion relation follows. 

In this paper we show that the ordinary analysis of QGP instabilities based on the Maxwell-Boltzmann equation may be reproduced in a purely (viscous) hydrodynamic model. To our knowledge, this is the first attempt of this kind that deals with a relativistic plasma and includes the effect of collisions. Manuel and Mr\'owczy\'nski have derived ideal hydrodynamic-like equations that are applicable to short-time scale color phenomena in the QGP and applied them to study the collective modes in a two-stream system \cite{manuel}. An alternative set of hydrodynamic equations for a non-Abelian plasma is derived in \cite{color}. 
Using the chromo-hydrodynamic formalism developed in \cite{manuel}, Mannarelli and Manuel investigated jet-induced stream instabilities in Ref. \cite{man1} and compared the results with those obtained with kinetic theory in Ref. \cite{man2}. 
Based on a particular closure and neglecting collisions, Basu has studied the Weibel instabilities in a non-relativistic plasma within a hydrodynamic model and has been able to reproduce, in the strong anisotropy limit, the results from kinetic theory \cite{basu}. 

For simplicity, we shall restrict ourselves to the instability of a homogeneous configuration of an EM plasma (i.e. the background EM field and the background one-particle distribution function are independent of spatial coordinates).  
We shall reproduce the kinetic theory analysis with a hydrodynamical effective theory derived from kinetic theory by use of the so-called Entropy Production Variational Method (EPVM) -- see Refs. \cite{epvm-rev,prigo} for a detailed account of this method, and Refs. \cite{jona,ruelle} for its connection to nonequilibrium statistical mechanics. We note that similar approaches to the EPVM have been applied to diverse transport phenomena in Refs. \cite{tanos1,tanos2,tanos3,christen,christen2}. 

As indicated above, we shall deal with an Abelian plasma. The extension to a non-Abelian plasma is left for the future, but it is worth mentioning that the developments presented here are relevant to the non-Abelian case under the Abelian dominance approximation valid for weak gauge fields $A \ll p_h$, where $A$ and $p_h$ are the vector potential amplitude and the characteristic momentum of hard particles (see, e.g., \cite{schenke-coll,rom3,dominik}). 

This paper is organized as follows. In Section \ref{setup} we give a brief overview of the kinetic theory and hydrodynamics of a charged fluid, and describe the closure obtained from the EPVM that we use to go from kinetic theory to the hydrodynamic effective theory on which our developments are based. In Section \ref{inst} we analyze the Weibel instability as obtained from the effective theory, study the dependence of its growth rate on the collision time and compare our results to those obtained from kinetic theory. Finally, in Section \ref{conc} we present our conclusions and outlook. 

\section{Theoretical setup}
\label{setup}

\subsection{Kinetic theory and hydrodynamics of a charged fluid}

Let us begin by briefly reviewing the kinetic theory and hydrodynamics of a charged fluid.

The fluid has two conserved quantities, the energy-momentum tensor $T^{\mu\nu}$ and current $J^\mu$. They obey conservation laws and the Maxwell equations

\bea 
T^{\mu\nu}_{,\nu}&=&F^{\mu\rho}J_{\rho}\nn
J^{\mu}_{,\mu}&=&0\nn
F^{\mu\nu}_{,\nu}&=&J^{\mu}
\tea 
where $F^{\mu\nu}$ is the strength tensor
\be 
F^{\mu\nu}=\left( 
\begin{array}{cccc}
0& E^1 & E^2 & E^3 \\
-E^1& 0 & B^3 & -B^2 \\
-E^2& -B^3 & 0 & B^1 \\
-E^3 & B^2 & -B^1 & 0
\end{array}
\right) 
\te 

For an ideal fluid (which is not the case studied here) we would have

\bea 
T^{\mu\nu}&=&\left(\rho+p \right)u^{\mu}u^{\nu}+pg^{\mu\nu}  \nn
J^{\mu}&=&qu^{\mu}
\tea
where $\rho$ is the energy density, $p$ is the pressure, and $q$ is the charge density.

In the kinetic theory description the transport equation reads 

\be
p^{\mu}\left[\partial_{\mu}f-eF_{\mu\nu}\frac{\partial f}{\partial p_{\nu}}\right]=\frac{-1}{\tau}\mathrm{sign}\left(p^0\right)I_{col}
\te
where $f=f(x^\mu,p^\mu,t)$ is the one-particle distribution function, $e$ and $p^\mu$ are the particles' charge and momentum, $I_{col}$ is the collision operator, and $\tau$ is the collision time. We shall specify $I_{col}$ later on.
The current is

\be
J^{\mu}=e\int\;Dp\;p^{\mu}f
\te
and the energy-momentum tensor is

\be
T^{\mu\nu}=\int\;Dp\;p^{\mu}p^{\nu}f
\te
where the measure is given by

\be
Dp =\frac{2d^4p\delta\left(p^2\right)}{\left(2\pi\right)^3}=\frac{d^4p}{\left(2\pi\right)^3 p}\left(\delta\left(p^0-p \right) + \delta\left(p^0+p \right)\right) 
\te
For simplicity we assume massless particles. 
To enforce the conservation laws we require

\be
\int\;Dp\;\mathrm{sign}\left(p^0\right)I_{col}=\int\;Dp\;\mathrm{sign}\left(p^0\right)p^{\mu}I_{col}=0
\te

We parametrize the distribution function as follows

\be 
f=f_B\left(1+Z\right)
\te 
with $Z=0$ at equilibrium. $f_B$ is the background solution. Following Weibel \cite{weibel}, we assume $f_B$ is independent of spatial coordinates. Later on we shall specify $f_B$, but the moment we leave this choice open.

Our developments will be based on the EPVM, so we shall make use of the expression giving the entropy production. 
We assume a simple Boltzmann type relative entropy flux \cite{vendra,peres,sagawa}

\be
S^{\mu}=-\int\;Dp\;\left(\mathrm{sign}\left(p^0 \right)\right)p^{\mu}\left[f\ln \left(\frac{f}{f_B}\right)-\left(f-f_B\right)\right]
\te
so we get the relative (with respect to $f_B$) entropy production

\be
S^{\mu}_{,\mu}=\frac{1}{\tau}\int\;Dp\;I_{col}\ln \left(1+Z\right)
\te

To continue we must make an ansatz regarding the collision operator. We shall assume a linear collision term of the form 

\be 
I_{col}\left( p\right) =\int\;Dp'\;K\left[ p,p'\right]  Z\left( p'\right) 
\label{lincoll}
\te
where $K$ is a symmetric operator. The idea behind Eq. (\ref{lincoll}) is that the antisymmetric part of the kernel of the linear collision operator does not affect the entropy production to lowest order. 
Retaining only the symmetric kernel in the linear collision term, we focus on the part of the dynamics which is directly related to the relaxation of the system towards homogeneity, since this relaxation entails an entropy increase.
The conservation laws then mean that $\mathrm{sign}\left(p^0\right)$ and $\mathrm{sign}\left(p^0\right)p^{\mu}$ are homogeneous solutions. This suggests writing

\be 
I_{col}\left( p\right) =F\left(p\right)f_B\left(p\right) \left\{Z\left( p\right)- \mathrm{sign}\left(p^0\right)\left[A[Z]+p^{\mu}B_{\mu}[Z]\right\}\right]
\te
where $F$ is some even function of $p^i$ ($i=1,2,3$), and the quantities $A[Z]$ and $B_{\mu}[Z]$ are functionals of $Z$ introduced to enforce the constraints, namely

\bea
\int\;Dp\;\left(A+B_{\nu}p^{\nu} \right) F  f_B&=& \int\;Dp\;\mathrm{sign}\left(p^0\right)Ff_BZ \nn
\int\;Dp\;p^{\mu}\left(A+B_{\nu}p^{\nu} \right)F  f_B&=&\int\;Dp\;\mathrm{sign}\left(p^0\right)p^{\mu}Ff_BZ
\tea

We shall not need the explicit expressions for $A[Z]$ and $B_{\mu}[Z]$, but they can be calculated following Ref.  \cite{color}. For the moment, we need not specify the function $F(p)$, but we note that $F=p$ corresponds to Anderson-Witting's ansatz for the collision term \cite{AW,taka}, which we shall use in the subsection \ref{anisub}.

\subsection{Closure from the EPVM}

In order to go from kinetic theory to a fluid description, we need to express $Z$ (the deviation from equilibrium) in terms of hydrodynamic variables, i.e., to provide a closure. The most well-known approaches are the Chapman-Enskog expansion and Grad's quadratic ansatz. As mentioned in the Introduction, in this paper we will use the closure that is obtained from the EPVM (a review can be found in Ref. \cite{epvm-rev}). We will now briefly review this formalism, focusing on the derivation of the hydrodynamic effective theory that will serve as our basis. A detailed account of this derivation can be found in Refs. \cite{linking,color}; see also \cite{net,app} for concrete applications to heavy ion collisions.  

Traditional fluid dynamics derived from kinetic theory by the Chapman-Enskog expansion or Grad's ansatz has two important limitations. First, it relies on an expansion in gradients of hydrodynamic variables, which necessarily implies that the system is sufficiently close to equilibrium so that these gradients are small. In turn, this means that the system is also very close to being isotropic in momentum space. Second, it breaks down at large shear viscosity $\eta$. The formalism obtained from the EPVM does not suffer from these drawbacks and can succesfully track the evolution as given by kinetic theory even for highly nonequilibrium flow.  By direct comparison to solutions to Boltzmann's equation, we have shown in \cite{net} that for the boost-invariant 1D expansion of matter created in heavy ion collisions the model obtained from the EPVM can reproduce the kinetic theory results, even for highly nonequilibrium situations and/or large values of the shear viscosity of the QGP. Instead, it is well-known that the full Israel-Stewart formalism \cite{hyd1,hyd2,hyd3} (based on Grad's ansatz) fails in these cases, because the pressure becomes negative due to the large values of the shear tensor (see, e.g., \cite{huov}).

If deviations from equilibrium are small, the EPVM closure reduces to Grad's quadratic ansatz \cite{linking}. 
The effective theory includes nonhydrodynamic variables in addition to the usual hydrodynamic ones (we shall call them $\zeta_{\lambda\sigma}$ and $\zeta_{\lambda}$ in what follows). These variables model the backreaction of $f$ --that may describe a highly nonequilibrium situation-- on the hydrodynamic modes which relax much more slowly. In other words, on time scales short with respect to $\tau$ the fluid relaxes to a steady nonequilibrium state characterized by a
nonvanishing viscous energy momentum tensor; the relaxation to true equilibrium is a much slower process.

In the EPVM, $\zeta_{\lambda\sigma}$ and $\zeta_{\lambda}$ are identified with the Lagrange multipliers of the variational problem whose solution gives the $f$ that extremizes the entropy production given fixed values of $T^{\mu\nu}$ and $J^\mu$. Physically, one can think of the EPVM as selecting the dynamics of these nonhydrodynamic variables in such a way as to extremize the production of entropy during the evolution of the system \cite{epvm-rev,prigo,linking}. 

The EPVM leads, to first order in Lagrange multipliers to the equation (see Refs. \cite{linking,color} for details)

\be 
2 I_{col}\left[ Z\right] =\tau f_B \left[\zeta_{\lambda\sigma}p^{\lambda}p^{\sigma}+\zeta_{\lambda}p^{\lambda} \right] 
\te 
Consistency requires 

\bea 
\int\;Dp\;f_B\mathrm{sign}\left(p^0 \right) \left[\zeta_{\lambda\sigma}p^{\lambda}p^{\sigma}+\zeta_{\lambda}p^{\lambda} \right] &=&0\nn
\int\;Dp\;f_B\mathrm{sign}\left(p^0 \right) p^{\mu}\left[\zeta_{\lambda\sigma}p^{\lambda}p^{\sigma}+\zeta_{\lambda}p^{\lambda} \right] &=&0
\tea 
We now get

\be 
Z=\frac{\tau}{2F}\left[\zeta_{\lambda\sigma}p^{\lambda}p^{\sigma}+\zeta_{\lambda}p^{\lambda} \right] +Z_{hom}
\te 
where

\be 
Z_{hom}=\mathrm{sign}\left(p^0\right)\left[\alpha +\beta_{\lambda}p^{\lambda}\right]
\te

Inserting this into the linearized transport equation we get

\be
p^{\mu}\left[f_B \partial_{\mu}\left\{\frac{\tau}{2F}\left[\zeta_{\lambda\sigma}p^{\lambda}p^{\sigma}+\zeta_{\lambda}p^{\lambda} \right] +\mathrm{sign}\left(p^0\right)\left[\alpha +\beta_{\lambda}p^{\lambda}\right]\right\}-eF_{\mu\nu}\frac{\partial f_B}{\partial p_{\nu}}\right] =\frac{-1}{2}\mathrm{sign}\left(p^0\right)f_B\left[\zeta_{\lambda\sigma}p^{\lambda}p^{\sigma}+\zeta_{\lambda}p^{\lambda} \right]
\te
The idea is to get equations for $\zeta_{\lambda\sigma}$, $\zeta_{\lambda}$, $\beta_{\lambda}$ and $\alpha$ by taking moments of this equation \cite{color,rischke,deneur}. We shall do this at linear order in the next section.

\section{Weibel instability}
\label{inst}

To focus on the physics of unstable modes, we assume $f_B=f\left(\vec{p}\right)\theta\left(p^0\right)$, where $f$ is even but anisotropic (as in Weibel's seminal paper \cite{weibel}). More concretely,
$f=C\left(\beta ,b\right)f_{\beta b}\left(p\right)$, where

\be
f_{\beta b}=e^{-\beta p}e^{-b^2p_z^2}
\te
as advocated in \cite{schlick}. $C(\beta,b)$ is a normalization factor to be fixed later on, and $b^2$ is a measure of the momentum anisotropy of the background configuration.

Note that considering the collisionless transport equation in the usual approach corresponds to the $\tau\to\infty$ limit. In the absense of a Maxwell field, in this limit only terms containing derivatives of the hydrodynamic fields remain in the equations of motion. In other words, any space-time independent configuration of fields $\alpha$, $\beta^{\mu}$, $\zeta^{\mu}$ and $\zeta^{\mu\nu}$ will be a solution of the hydrodynamic equations, provided the EM current vanishes (since we want to avoid having a background EM field). To achieve this, we assume that there is a compensating static background.  
On the other hand, it makes no sense to linearize in $\tau$ (this rules out the possibility of using the closure provided by Grad's ansatz or the Chapmann-Enskog expansion). For this reason, we shall not write down the full equations of motion, but derive the linearized equations directly from kinetic theory instead.

To be even more concrete and to simplify matters, we seek solutions with ($a=1,2$)
$\zeta_{00}=\zeta_{ab}=\zeta_{0a}=\zeta_{03}=\zeta_{33}=0$, $\zeta_{\lambda}=\alpha=0$, $\beta_0=\beta_3=0$. We further assume a plane wave solution propagating in the $z$ direction, namely the space time dependence of all factors is of the form $\exp\left\{i\left(kz-\omega t\right)\right\}$. The consistency conditions hold, and the transport equation becomes

\be
if_{\beta b} \left(kp^3-\omega p\right)\left\{\frac{\tau}{F}\zeta_{b3}p^{b}p^{3}  +\beta_{b}p^{b}\right\}-ep^{\mu}F_{\mu\nu}\frac{\partial f_{\beta b}}{\partial p_{\nu}} =-f_{\beta b}\zeta_{b3}p^{b}p^{3} 
\te
Observe that the normalization drops out.

To compute the moments of this equation, observe that

\bea
\int\;Dp\;p^{\mu}F_{\mu\nu}\frac{\partial f_{\beta b}}{\partial p_{\nu}}&=&0\nn
\int\;Dp\;p_{\lambda}p^{\mu}F_{\mu\nu}\frac{\partial f_{\beta b}}{\partial p_{\nu}}&=&-\int\;Dp\;p^{\mu}F_{\mu\lambda}
 f_{\beta b}\nn
\int\;Dp\;p_{\lambda}p_{\rho}p^{\mu}F_{\mu\nu}\frac{\partial f_{\beta b}}{\partial p_{\nu}}&=&-\int\;Dp\;p^{\mu}\left[F_{\mu\lambda}p_{\rho}+F_{\mu\rho}p_{\lambda}\right]
 f_{\beta b}
\tea
We introduce the notation

\be
\left\langle A\right\rangle=\int\;\frac{d^3p}{\left(2\pi\right)^3 p}f_{\beta b} A
\te
The nontrivial first order moments of the kinetic equation are

\bea
ik\tau\zeta_{b3}\left\langle \frac1Fp^ap^bp_3^2\right\rangle -i\omega\beta_b\left\langle pp^ap^b\right\rangle+eF_{0a}\left\langle p\right\rangle&=&0\nn
F_{03}\left\langle p\right\rangle&=&0\nn
\tea
The nontrivial second order moments are

\bea
ik\tau\zeta_{b3}\left\langle \frac1Fpp^ap^bp_3^2\right\rangle -i\omega\beta_b\left\langle p^2p^ap^b\right\rangle+e\left[F_{b0}\left\langle p^ap^b\right\rangle+F_{0a}\left\langle p^2\right\rangle\right]&=&0\nn
ik\beta_b\left\langle p^ap^bp_3^2\right\rangle-i\omega\tau\zeta_{b3}\left\langle \frac1Fpp^ap^bp_3^2\right\rangle+e\left[F_{3a}\left\langle p_3^2\right\rangle+F_{b3}\left\langle p^ap^b\right\rangle\right]&=&-\zeta_{b3}\left\langle p^ap^{b}p_{3}^2 \right\rangle\nn
\tea
We shall only use the second of these equations. Finally the current is

\be
J^{\mu}=eC\left(\beta ,b\right)\left\langle p^{\mu}\left[\frac{\tau}{F}\zeta_{b3}p^{b}p^{3} +\beta_{b}p^{b}\right]\right\rangle
\te
The only nonzero component of the current is

\be
J^{a}=eC\left(\beta ,b\right)\beta_{b}\left\langle p^{a}p^{b}\right\rangle
\te

So, the nontrivial Maxwell equations are

\be
i\omega F_{a0}+ikF_{a3}=J_a=eC\left(\beta ,b\right)\beta_{b}\left\langle p_{a}p^{b}\right\rangle
\te
and 

\be
-i\omega F_{a3}-ikF_{a0}=0
\te

To write the final forms of the equations we observe that

\be
\left\langle p^ap^bA\left(p,p^3\right)\right\rangle=\frac12\delta^{ab}\left\langle \left(p^2-p_3^2\right)A\left(p,p^3\right)\right\rangle
\te


Let us call

\bea
\left\langle p\right\rangle&=& \frac nC\nn
\frac12 C\left\langle \left(p^2-p_3^2\right)\right\rangle&=&p_x\nn
\left\langle p^2-3p_3^2\right\rangle&=&\frac 2C\left(p_x-p_z\right)
\tea
and

\bea
\left\langle p\left(p^2-p_3^2\right)\right\rangle&=& A_3\nn
\left\langle \left(p^2-p_3^2\right)p_3^2\right\rangle&=& A_4\nn
\left\langle \frac1F\left(p^2-p_3^2\right)p_3^2\right\rangle&=& A_{4F}\nn
\left\langle \frac1Fp\left(p^2-p_3^2\right)p_3^2\right\rangle&=& A_{5F}
\tea
Therefore

\be
\left(
\begin{array}{cccc}
\omega     & k      & iep_x      & 0 \\
k          & \omega &  0           & 0 \\
-i2e\frac nC & 0                                   & \omega A_3 & -k\tau A_{4F} \\
0           & \frac {2ie}C\left(p_x-p_z\right) & - k A_4      & iA_4+\omega\tau A_{5F}
\end{array}
\right)\left(
\begin{array}{c}
F_{a0}\\
F_{a3}\\
\beta_a\\
\zeta_{a3}
\end{array}
\right)=0
\te

If $k=0$, the dispersion relation is

\be
\left[iA_4+\omega\tau A_{5F}\right]\omega \left[A_3\omega^2- 2e^2\frac nCp_x\right]=0
\te
Thus we identify the plasma frequency

\be
\omega_p^2=\frac {2e^2n}{CA_3}p_x
\te

In general, the dispersion relation is

\be
\left[iA_4+\omega\tau A_{5F}\right]A_3\omega\left[\omega^2-k^2-\omega_p^2\right]-k^2\tau A_{4F}\left[ A_4\left(\omega^2-k^2\right)+\frac {2e^2p_x}C\left(p_x-p_z\right)\right]=0
\label{general}
\te
Let us call

\be
k_{st}^2=\frac {2e^2p_x}{CA_4}\left(p_x-p_z\right)
\te
Note that $k_{st}$ is a measure of the anisotropy of the background distribution function $f_B$.




Defining

\be
\tau_0=\frac{A_4}{A_{5F}}
\te

\be
\lambda =\frac{A_{4F}A_4}{A_3A_{5F}}
\te


and putting 

\be
\omega =i\sigma
\te
the dispersion relation becomes 

\be
P\left[ \sigma\right] = \sigma^4+\frac {\tau_0}{\tau}\sigma^3+\left\{k^2+\omega_p^2+\lambda k^2 \right\}\sigma^2+\frac {\tau_0}{\tau}\left[k^2+\omega_p^2\right]\sigma-\lambda k^2 \left\{k_{st}^2-k^2\right\}=0
\te
It is clear that $P\left[ \sigma\right] $ increases with $\sigma$ when $\sigma$ is real and positive. Therefore, for $0<k^2<k_{st}^2$ there is one (and only one) real positive root. This root corresponds to the Weibel instability, on which we focus in what follows.

\subsection{Collision time dependence}
To investigate how the growth rate of the Weibel instability depends on $\tau$, observe that if $\sigma\left( \tau\right) $ denotes the root for a given $\tau$, then

\be 
P'\left[\sigma\left( \tau\right) \right] \frac{d\sigma}{d\tau}-\frac{\tau_0}{\tau^2}P_2\left[\sigma\left( \tau\right) \right]=0
\te 
where

\be 
P_2\left[\sigma \right]=\sigma\left[\sigma^3+k^2+\omega_p^2 \right] 
\te 
Since $P'$ and $P_2$ are positive for positive $\sigma$, we must have $d\sigma/d\tau\ge 0$. Also write

\be
P\left[ \sigma\right] = P_1\left[\sigma \right]+\frac{\tau_0}{\tau}P_2\left[\sigma \right]
\te 
 
\be 
P_1\left[\sigma \right]= \sigma^4+\left\{k^2+\omega_p^2+\lambda k^2 \right\}\sigma^2-\lambda k^2 \left\{k_{st}^2-k^2\right\}=0
\te
Then we must have 

\be 
P_1\left[\sigma\left( \tau\right) \right]\le 0
\te 
which implies $\sigma\left( \tau\right)\le\sigma_{max}$, where

\be 
\sigma_{max}^2=\frac12\left[ \sqrt{4\lambda k^2 \left\{k_{st}^2-k^2\right\}+\left\{k^2+\omega_p^2+\lambda k^2 \right\}^2}-\left\{k^2+\omega_p^2+\lambda k^2 \right\}\right] 
\te 
Since $\sigma_{max}^2=0$ both for $k^2=0$ and $k^2=k_{st}^2$, it must have a maximum in $k$. The maximum is achieved when

\be 
\left( 1+\lambda\right) \sigma_{max}^2-\lambda k_{st}^2+2\lambda k^2=0 
\te 
We have $\sigma\left(\tau \right)\to\sigma_{max}$ for $\tau\to\infty$, which corresponds to the collisionless limit usually assumed to study the Weibel instability. This means that, as expected on physical grounds, collisions slow down the growth of the unstable mode. The particles with momentum orthogonal to the magnetic field, which are responsible for the Weibel instability, can scatter with other particles and avoid getting trapped in the background magnetic field, so a smaller current and therefore a smaller induced magnetic field are generated. This has been verified directly from kinetic theory with a BGK collision term \cite{schenke-coll}. 

For $\tau\to 0$, corresponding to the ideal fluid limit, we get instead  

\be 
\sigma \left(\tau \right) =\frac{\tau}{\tau_0}\frac{\lambda k^2}{k^2+\omega_p^2}\left(k_{st}^2-k^2 \right)
\label{tauto0}
\te 
In this regime, the maximum growth rate is achieved at

\be 
k^2=\omega_p^2\left[ \sqrt{1+\frac{k_{st}^2}{\omega_p^2}}-1\right] 
\te 
and takes the value

\be 
\sigma \left(\tau \right) =\frac{\tau}{\tau_0}\lambda \omega_p^2
\left[ \sqrt{1+\frac{k_{st}^2}{\omega_p^2}}-1\right] ^2
\label{diss}
\te 
This shows that the instability dissappears for all practical purposes when $\tau$ is small enough, because there is a maximum time scale over which the theory makes sense given by the time scale on which the background solution relaxes to equilibrium. This feature is also present in the kinetic theory approach of Ref. \cite{schenke-coll}. 

\subsection{Anisotropy dependence}
\label{anisub}
We wish to discuss now how the instability growth rate depends on the parameters $\beta$ and $b$. In order to obtain analytic results, we shall restrict our discussion to the strong anisotropy limit $b \rightarrow \infty$. We use the Anderson-Witting collision term \cite{AW,taka}, that corresponds to $F=p$. 

Let us define

\be
\left\langle 1\right\rangle=\int\;\frac{d^3p}{\left(2\pi\right)^3 }\;p^{-1}e^{-\beta p}e^{-bp_z^2}=\frac1{\left(2\pi\right)^2 }\int_0^{\infty}p\;dp\int_0^{\pi}\sin\theta\;d\theta\;e^{-\beta p}e^{-b^2p^2\cos^2\theta}
\te

The idea is that all the necessary expectation values may be obtained as derivatives of $\left\langle 1\right\rangle $ according to the rule

\be 
\left\langle p^A\left( p_z^2\right) ^B\right\rangle =\left( -1\right) ^{A+B}\frac{\partial^{A+B}}{\partial\beta^A\partial \left( b^2\right)^B }\left\langle 1\right\rangle
\te 
Note, however, that for the Anderson-Witting collision term \cite{AW,taka} $A_{4F}$ cannot be obtained this way, but rather from the solution to the equation

\be 
-\frac{\partial}{\partial\beta}A_{4F}=A_4
\te 

Call $q=b/\beta$

\be
\left\langle 1\right\rangle=\frac 1{2\pi^2\beta^2 }\int_0^{\infty}t\;dt\int_0^{1}dx\;e^{-t}e^{-q^2t^2x^2}
\te
Following \cite{tabla}, we define the error function

\be
\Phi\left[z\right]=\mathrm{erf}\left[z\right]=\frac 2{\sqrt{\pi}}\int_0^z\;dt\;e^{-t^2}
\te
so we get 



\be
\left\langle 1\right\rangle=\frac 1{4\pi^{3/2}\beta b }\left[1-\Phi\left[\frac 1{2q}\right]\right]e^{1/4q^2}
\label{1exact}
\te
Note that no approximation regarding the value of $b$ has been done to get Eq. (\ref{1exact}).

In the strong anisotropy limit $b\to\infty$, (\ref{1exact}) reduces to

\be
\left\langle 1\right\rangle_{\infty}=\frac 1{4\pi^{3/2}\beta \left( b^2\right)^{1/2} }
\te
In this regime we obtain (in all cases to leading order in $b^{-1}$):

\bea
\left\langle p\right\rangle &=&\frac nC=\frac 1{4\pi^{3/2}}\frac{1}{\beta^2b}\nn 
\left\langle p^2\right\rangle &=&\frac {2p_x}C=\frac 1{4\pi^{3/2}}\frac{2}{\beta^3b}\nn 
\left\langle p^3\right\rangle &=&A_3=\frac 1{4\pi^{3/2}}\frac{6}{\beta^4b}\nn 
\left\langle p^2p_z^2\right\rangle &=&A_4=A_{5F}=\frac 1{4\pi^{3/2}}\frac{1}{\beta^3b^3}\nn 
\left\langle pp_z^2\right\rangle &=&A_{4F}=\frac 1{4\pi^{3/2}}\frac{1}{2\beta^2b^3}
\tea 

We wish to consider configurations with different anisotropy parameter $\xi=b^2/\beta^2$ but the same $\beta$ and particle density $n=n_0$. Thus we define 

\be 
C=\frac{n_0}{\left\langle p\right\rangle } =4\pi^{3/2}n_0\beta^2b
\te 
In this case 

\be 
\omega_p^2=\frac13e^2n_0\beta 
\te 
and $\tau_0=1$. The quantity $k_{st}^2$ is unbounded

\be 
k_{st}^2=2\left( n_0\beta^3\right)\frac{b^2}{\beta^4} 
\te 
Instead $\lambda$ goes to zero

\be 
\lambda =\frac{\beta^2}{12b^2}
\te 
Note that the product $\lambda k_{st}^2=n_0\beta /6$ remains bounded. If $\tau$ is small (but not so small as to make the Weibel instability dissappear --see the discussion after Eq. (\ref{diss})), we have from Eq. (\ref{tauto0}) that 
\be 
\sigma(\tau) = \tau [\lambda k_{st}^2]\frac{k^2}{k^2+w_p^2}
\label{ksat}
\te 
The behavior of $\sigma$ with increasing $k$ that we obtain agrees well with the kinetic theory results of \cite{schenke-coll} (see Fig. 11 of that reference). The growth rate increases with increasing $k$ as given by Eq. (\ref{ksat}) and for large $k$ it saturates at $\tau [\lambda k_{st}^2]$. 

\section{Conclusions}
\label{conc}

We have studied the Weibel instability within the framework of an hydrodynamic effective theory derived by the Entropy Production Variational Method from the relativistic Maxwell-Boltzmann equation. This method provides a fluid closure that yields an effective theory capable of describing highly nonequilibrium flows. 

We have found that the usual linear analysis of QGP Weibel instabilities based on the Maxwell-Boltzmann equation may be reproduced in this particular hydrodynamic model. We have analyzed the dependence of the growth rate of the Weibel instability on the collision time and found that, as expected, the effect of collisions is to slow down the growth of the magnetic field's amplitude. We have also shown that if the collision time is too short, corresponding to a nearly perfect fluid, the Weibel instability dissapears. Our results agree with those obtained from kinetic theory with a  BGK collision term \cite{schenke-coll}. 

We believe that this study opens up the possibility of investigating the effect of non-linear self-interactions of the gauge fields on the chromo-Weibel instability relevant to QGP dynamics, within a framework that is simpler than the kinetic theory of non-Abelian plasmas including particle collisions. This may be useful to investigate some issues which are hard to attack from kinetic theory, such as the backreaction of Weibel instabilities on the dynamics of the rapidly expanding QGP \cite{attems1,attems2,rom3,ven}. To this end, we must rely on the extension to non-Abelian plasmas of the effective hydrodynamic formalism presented here, which we have carried out in \cite{color}, and study the development of the chromo-Weibel instability in this setting. Work is in progress along this line.  

\begin{acknowledgements}
This work has been supported in part by ANPCyT, CONICET and UBA under Project UBACYT X032 (Argentina).
 \end{acknowledgements}

\end{document}